\documentclass{JHEP3}
\usepackage{epsfig,multicol,bbm}
\usepackage{amssymb,amsmath}
\usepackage{latexsym}
\usepackage{graphicx}
\usepackage{epsfig}
\usepackage{epsf}
\usepackage{cite}
\usepackage{rotating}
\usepackage{graphicx}
\usepackage{latexsym}
\usepackage{amsfonts}
\usepackage{epic}
\usepackage{lscape,fancyhdr,array}

\def\eps{\epsilon}

\def\d{\partial}
\def\cA{{\cal A}}

\def\cH{{\cal H}}
\def\cI{{\cal I}}

\def\cK{{\cal K}}
\def\cL{{\cal L}}

\def\cN{{\cal N}}

\def\cQ{{\cal Q}}

\def \p{{\partial}}

\def \eps{{\epsilon}}

\newcommand{\re}[1]{(\ref{#1})}

\def\oneone{\rlap 1\mkern4mu{\rm l}}

\def\buildrel#1_#2^#3{\mathrel{\mathop{\kern 0pt#1}\limits_{#2}^{#3}}}

\newcommand{\ffrac}[2]{\raisebox{.5pt}
  {\footnotesize $\displaystyle\frac{#1}{#2}$}\kern1pt}

\newcommand{\ls}{{\frak{sl}(2,\mathbb{R})}{}}

\newcommand{\sch}{Schr\"odinger }

\renewcommand{\d}{\partial}

\def\buildrel#1_#2^#3{\mathrel{\mathop{\kern 0pt#1}\limits_{#2}^{#3}}}
\newcommand{\half}{\mathchoice{%
    \ffrac{1}{2}}{\frac{1}{2}}{\frac{1}{2}}{\frac{1}{2}}}

\def\no{\nonumber}
\def\beq{\begin{eqnarray}}
\def\eeq{\end{eqnarray}}

\def\ie{{\it i.e.~}}

\title{Asymptotic symmetries of Schr\"odinger spacetimes}

\author{Geoffrey Comp\`ere$^{\flat}$, Sophie de Buyl$^{\flat}$, St\'ephane Detournay$^{\natural}$, and Kentaroh Yoshida$^{\natural,\dagger}$
~\\
{$^{\flat}$  Department of Physics, 
University of California, Santa Barbara, 
Santa Barbara, CA 93106, USA
 }
 ~\\
 {$^{\natural}$ \it Kavli Institute for Theoretical Physics, University of California, Santa Barbara, CA 93106, USA
}
~\\
{$^{\dagger}$ \it Department of Physics, Kyoto University, Kyoto 606-8502, Japan
}
~\\
~\\
{E-mail: {\tt gcompere@physics.ucsb.edu, sdebuyl@physics.ucsb.edu, detourn@kitp.ucsb.edu, kyoshida@gauge.scphys.kyoto-u.ac.jp}}
 }
\received{\today} 		%%
\preprint{arXiv:0908.1402}	% OR: \preprint{Aaaa/Mm/Yy\\Aaa-aa/Nnnnnn}
\abstract{We discuss the asymptotic symmetry algebra of the Schr\"odinger
-invariant metrics in $d+3$ dimensions and its realization on finite temperature solutions of gravity coupled to matter fields. These solutions have been proposed as gravity backgrounds dual to non-relativistic CFTs with critical exponent $z$ in $d$ space dimensions. It is known that the Schr\"odinger
algebra possesses an infinite-dimensional extension, the Schr\"odinger-Virasoro algebra. However, we show that the asymptotic symmetry algebra of Schr\"odinger spacetimes is only isomorphic to the exact symmetry group of the background. It is possible to construct from first principles finite and integrable charges that infinite-dimensionally extend the Schr\"odinger algebra but these charges are not correctly represented via a Dirac bracket. We briefly comment on the extension of our analysis to spacetimes with Lifshitz symmetry.}

\keywords{Gauge-gravity correspondence, Non-relativistic CFT, asymptotic symmetries}

\begin{document}

\section{Introduction}

The holographic description \cite{Maldacena:1997re,Gubser:1998bc,Witten:1998qj} of 
condensed matter systems such as superconductors \cite{Gubser:2008px,Hartnoll:2008vx} and materials undergoing the quantum Hall effect \cite{KeskiVakkuri:2008eb,Davis:2008nv,Fujita:2009kw} have recently attracted a lot of interest.
The systems are strongly coupled at critical points and hence 
the holographic description may give a new analytical method 
to investigate some aspects of the critical behaviors 
in terms of classical gravity. 

\medskip 

Some condensed matter systems realized in laboratories are described at their critical points by non-relativistic 
conformal field theories (NRCFTs). Non-relativistic conformal symmetry contains the scaling invariance 
\[
t \to \lambda^z t\,, \qquad x^i \to \lambda x^i\,,
\]
where $z$ is a dynamical exponent. When $z=2$, the symmetry is enhanced to the Schr\"odinger symmetry \cite{Hagen:1972pd,Niederer:1972zz} containing in addition the special conformal transformations. The NRCFTs based on the \sch symmetry are studied e.g. in \cite{Henkel:1993sg, Mehen:1999nd, Son:2005rv,Nishida:2007pj,Bobev:2009mw,Donos:2009xc}.

\medskip

Recently, gravity duals for these NRCFTs have been proposed \cite{Son:2008ye,Balasubramanian:2008dm} (see \cite{Duval:1990hj}  for earlier work on the geometric realization of the Schr\"odinger symmetry and \cite{Duval:2008jg} for the relationship with \cite{Son:2008ye,Balasubramanian:2008dm}). The background at zero temperature consists in a light-like deformation of the anti-de Sitter metric -- for other gravity solutions and their string theory embedding, see \cite{Goldberger:2008vg,Barbon:2008bg,Herzog:2008wg,Maldacena:2008wh,Adams:2008wt,Kovtun:2008qy,Hartnoll:2008rs,Schvellinger:2008bf,Mazzucato:2008tr,Rangamani:2008gi,Adams:2008zk,Donos:2009en,Colgain:2009wm,Ooguri:2009cv,Donos:2009xc,Donos:2009zf}. The background metric is given by
\begin{equation}
\label{Metric}
ds^2 = L^2 \left(  \frac{dx^id x^i + 2 dx^+ dx^-}{r^2} 
+ \frac{dr^2}{r^2} \mp \frac{(dx^-)^2}{r^{2z}} 
\right)\,,
\end{equation}
where the $x^+$ direction is compactified as $x^+ \sim x^+ + 2\pi x^+_0$\, for some $x_0^+$. 
Both the deformation term and the compactification break the relativistic conformal symmetry to the \sch symmetry $\mathfrak{sch}_z(d)$ where $d$ is the number of space dimensions of the NRCFT. The isometries of this metric are identified with the time translations $H$, dilations $D$, mass/particle number $N$, spatial translations $P_i$, Galilean boosts $K_i$, and spatial rotations $M_{ij}$, with $i,j=1,...,d$. For $z=2$, an additional generator is present, corresponding to special conformal transformations $C$, which together with $H$ and $D$ form an $\ls$ subalgebra. Note that in the NRCFT context the minus sign should be taken in \re{Metric} so that the causality properties of a non-relativistic system will be recovered close to the boundary, see e.g. \cite{Brecher:2002bw,Hubeny:2002zr} and \cite{Hartnoll:2009sz}. The plus sign turns out to be relevant to describe black holes in three dimensions (i.e. $d=0$ above) \cite{Anninos:2008fx}.
\medskip 

The infinite-dimensional extension of the $\ls \oplus \ls$ symmetry algebra to two copies of the Virasoro algebras around $AdS_3$ in Einstein gravity \cite{Brown:1986nw} leads to severe constraints on the quantum theories dual to asymptotically $AdS_3$ spacetimes \cite{Strominger:1997eq,Maldacena:1998bw}. Now, it has been known for a while that an infinite-dimensional extension is possible for the \sch algebra with $z=2$ {\itshape in any dimension} (note this is also true for the isometry group $so(2,d-1)$ of $AdS_d$, given by the affine extension $\widehat{so(2,d-1)}$; the latter is however not realized as asymptotic symmetry algebra of $AdS_d$ \cite{Henneaux:1985tv}) . It is called the Schr\"odinger-Virasoro algebra \cite{Henkel:1993sg}. Moreover, such an algebra can be easily generalized to extend the $\mathfrak{sch}_z(d)$ symmetry for any $z$. The $\ls$ part of the \sch algebra has the familiar Virasoro extension while the other generators get enhanced to current algebras. As we will show below and as was shown independently in \cite{Alishahiha:2009nm}, one can represent the entire Schr\"odinger-Virasoro algebra as generators acting as diffeomorphisms on the solution \eqref{Metric}. All these generators are thus \emph{candidates} to be asymptotic symmetries of an ought-to-be defined phase space. One would like however to go beyond a kinetic analysis and learn if part of these symmetries can be dynamically realized around the gravity backgrounds \eqref{Metric} by attempting to construct a phase space accommodating these asymptotic symmetries. This is the main aim of the paper.

\medskip

More precisely, we will address the following issues
\begin{itemize}
\item Could the charges associated with the Schr\"odinger-Virasoro algebra be defined ? 

\medskip
One first observation one can make from the outset from purely algebraic considerations is the following. The author of \cite{Unterberger:2007hd} classified all possible central extensions for the Schrodinger-Virasoro algebra, showing that only the Virasoro could be centrally extended.  This means in particular that the current algebra with generators $N_m$ whose zero mode is the number operator (see \re{exact1}) has level $k=0$. But it is known that the only unitary representation of such an algebra is the trivial one, for which $N_m=0, \, \forall m$. Hence, the corresponding gravity dual would only be able to describe field theories with zero number operator, which would be of little interest!\footnote{We thank P. Ho\v{r}ava and Ch. Melby-Thompson for sharing that observation with us.}  However, non-unitary representations are not excluded on general grounds, since little is known on the nature of the field theory dual. On the other hand, infinite-dimensional extensions are possible even if the current algebra is not realized.

\item Could the charges represent the Schr\"odinger-Virasoro algebra ? 
\end{itemize}

Since the Schr\"odinger spacetimes are not asymptotically AdS, the holographic renormalization techniques based on Fefferman-Graham expansions used extensively in the AdS/CFT correspondence to compute the charges \cite{Balasubramanian:1999re,de Haro:2000xn} are not directly applicable but can be used as a guideline for extrapolating the charges, see e.g. the discussion in Appendix C of \cite{Martelli:2009uc}. While conserved charges for black holes can be defined using the regulated on-shell action for a phase space with fixed temperature and chemical potential \cite{Herzog:2008wg}, a Hamiltonian \cite{Regge:1974zd,Brown:1986ed} or Lagrangian \cite{Barnich:2001jy,Barnich:2003xg,Barnich:2007bf} definition of conserved charges is necessary in order to obtain a representation of the Schr\"odinger symmetries via a Dirac bracket. The conserved charges obtained via holographic and Hamiltonian/Lagrangian methods are identical up to background shifts at least for AdS spacetimes, see e.g. \cite{Hollands:2005wt,Papadimitriou:2005ii}. Note also that a holographic stress-tensor for Schr\"odinger space-times has recently been defined \cite{Ross:2009ar}.

The Lagrangian methods \cite{Barnich:2001jy,Barnich:2003xg,Barnich:2007bf} can be used straightforwardly, see Appendix \ref{method} for a short summary. Note however that in general the asymptotic charges of \cite{Barnich:2001jy,Barnich:2003xg,Barnich:2007bf} could be corrected due to counterterms in the regulated action \cite{Compere:2008us}, see \cite{Azeyanagi:2009wf,Amsel:2009pu} for related discussions. Counterterms can be obtained for a phase space of fluctuations around a fixed Schr\"odinger black brane \cite{Herzog:2008wg} and since they do not contain derivatives of the fields, they do not contribute to the charges. The computation of charges of \cite{Barnich:2001jy,Barnich:2003xg,Barnich:2007bf} however requires a phase space containing also the zero temperature background and counterterms for such general variations are unfortunately very difficult to obtain due to non-linearities at infinity (see however \cite{Ross:2009ar} for progress on this issue). We will assume in what follows that the supplementary counterterms, if any, to those of \cite{Herzog:2008wg} do not contribute to the definition of covariant phase space charges.

One difficulty than one faces right away is that the charges should be defined as an integral over $x^+$ and the infinitely extended spatial directions $x^i$. A regulator along the spatial directions is therefore needed in order to define finite charges. The approach taken in this paper is to consider that the $x^i$ have a finite extent, i.e. that the NRCFT is defined in a ``box''. We will see that the introduction of this regulator is sufficient to be able to define the charges associated to the Schr\"odinger-Virasoro group. 

It has been proven some time ago that once canonical charges associated with the asymptotic symmetries are defined and once a phase space preserved by the symmetries is shown to exist, the charges represent the algebra of asymptotic symmetries through a Dirac bracket \cite{Brown:1986ed}, see also \cite{Barnich:2007bf} for the analogous theorem in Lagrangian formalism. Here, however, one cannot use blindly these theorems since the regulator may also be transformed under the asymptotic symmetries. Only a subset of the proposed Schr\"odinger-Virasoro generators will preserve the regulator and will thus be represented under the usual bracket. We will propose a modified bracket which includes a change of regulator in order to treat the other candidate symmetries. The status of these transformations will be commented in the conclusions.

Spacetimes admiting the Lifshitz symmetry have also been considered recently in the holographic context, see \cite{Kachru:2008yh}. We will also discuss briefly in Appendix \ref{Lifsec} the extension of our analysis to those spacetimes.

The organization of this paper is as follows. The asymptotic analysis is presented in section \ref{dbigger0}. The method used is illustrated in some detail. We will mainly discuss the dynamical exponent $z = 2$ in all dimensions 
 and extend the analysis to an example with dynamical exponent bigger than 2, 
 namely $z = 3$ (for $d=2$). We finally conclude and interpret our results in section \ref{conclusion}. A short review of the method to compute (asymptotic) charges in the covariant formalism is given in Appendix \ref{method}, while asymptotic symmetries of Lifshitz spacetimes are discussed in Appendix \ref{Lifsec}.

\section{Asymptotic analysis of gravity duals to NRCFTs  \label{dbigger0}}

This section is devoted to a detailed analysis of the asymptotic symmetries of the Schr\"odinger metrics for $d>0$ and their realization via an algebra of asymptotic charges on some class of metrics that asymptotes to it. The backgrounds \re{Metric} possess $d$ space directions and are part of the ten or eleven dimensional metrics conjectured to be dual to $d$ dimensional non-relativistic systems.

\medskip

\noindent The successive steps to define an asymptotic algebra of charges that we will implement in the sequel are the following:

\begin{enumerate}
\item We will start by defining a class $\mathcal C$ of candidate asymptotic Killing vectors of the metrics \re{Metric} by solving the Killing equations up to a well chosen order in the $r$ expansion led by intuition. Solving the Killing equations to all orders in $r$ would result in finding only the exact Killing vectors, while solving only at the leading order would lead to a very large set of candidate asymptotic symmetries\footnote{See \cite{Compere:2007az}, page 142 and the appendix of \cite{Barnich:2006av} for a detailed  explanation on how to solve the Killing equations at first order in the radial expansion. The resolution at a given subleading order is then straightforward.}.
\item 
We will then construct a phase space ${\cal F}$ together with a class of asymptotic symmetries $\mathcal A \subset \mathcal C$ satisfying the following conditions : 
\begin{enumerate} 
\item The metrics in ${\cal F}$ should approach \re{Metric} in the limit $r \rightarrow 0$. 
\item The phase space ${\cal F}$ must contain solutions of interest such as black hole solutions. 
\item ${\cal F}$ must be invariant under the action of the finite diffeomorphisms associated with asymptotic Killing symmetries belonging to $\mathcal A$.
\item The asymptotic charges of the metrics in ${\cal F}$ associated with elements of $\mathcal A$ must be finite, conserved, and integrable. The asymptotic charges are computed by using the methods of \cite{Barnich:2001jy,Barnich:2003xg,Barnich:2007bf} which are briefly reviewed in Appendix \ref{method}. 
\end{enumerate}
The phase space is usually specified by a set of boundary conditions. Here instead, we will give an explicit construction of ${\cal F}$ and $\mathcal A$ starting from the candidate asymptotic Killing vectors. Only a subset of the candidate asymptotic Killing vectors $\mathcal C$ will be promoted to asymptotic symmetries $\mathcal A$ of ${\cal F}$. 

\item In the phase space ${\cal F}$, we will then study the algebra of asymptotic charges which should be isomorphic to the algebra of asymptotic symmetries  up to possible central extensions. 

\end{enumerate}

We start by deriving candidate asymptotic Killing vectors by solving the asymptotic Killing equations in section \ref{candasym}. We then turn to the realization of the candidate asymptotic symmetries on a phase space. Black hole solutions that asymptote the metrics \re{Metric}  are not known for general dimensions and critical exponents $z$. Therefore, we will first focus in section \ref{knsection} on the exponent $z=2$ for which $d$-dimensional black hole solutions \cite{Kovtun:2008qy}, generalizing the ones of \cite{Herzog:2008wg,Adams:2008wt,Maldacena:2008wh}, are known. In section \ref{MBH}, we treat the exponent $z=3$ in five dimensions $D=5$ ($d=2$) using solutions obtained by acting with the Null Melvin Twist on non-extremal D3-brane solutions \cite{Hubeny:2005qu} as an example of critical exponent greater than 2.

\subsection{Candidate asymptotic Killing vectors \label{candasym}}

For $z > 1$, the term $\frac{1}{r^{2z}}(dx^-)^2$ is the leading divergent term close to the boundary. This asymptotic behavior differs from asymptotically flat or anti-de Sitter spacetimes. When solving the Killing equations
\begin{equation}
\cL_{\xi_{as}} g_{\mu \nu} \rightarrow 0 \hspace{1cm} \mbox{for} \, r \rightarrow 0 , \end{equation}
up to certain well chosen orders (depending on each $\mu \nu$ component), we obtain the following vector fields, 
\begin{eqnarray}
 \xi_{as} &=& \frac{r}{z} L'(x^-) \p_r  + L(x^-) \p_{-} \nonumber \\ 
&& + \left( N(x^-)-\frac{z-2}{z}\,x^+\, L'(x^-) - \vec{x} \cdot \vec{X}'(x^-) -\frac{\vec{x}^2 + r^2}{2z} L''(x^-)\right) \p_+ \nonumber \\ 
&& + \left(X_i(x^-) +\frac{x_i}{z} L'(x^-) +  M_{ij} x_j \right) \d_i\, ,\label{candidakv}
\end{eqnarray}
where $M_{ij}$ is antisymmetric. The exact Killing vectors are recovered when $L''(x^-) = 0$, $N'(x^-) = 0$ and $X_i''(x^-) = 0$.  A detailed analysis implies that the rotations cannot be extended to $x^-$-dependent functions.

Defining the generators
\begin{eqnarray}
\hat L_n &=& \xi(L(x^-)= -2^{-n/2}(x^-)^{n+1})  \, \,  \, \, \, \text{for }n \in \mathbb Z \nonumber ,\\
\hat N_n &=& \xi(N(x^-) = 2^{-n/2}(x^-)^n)  \qquad \, \, \, \,  \, \, \, \text{for }n \in \mathbb Z \label{modesLn},\\
\hat X^i_{n} &=& \xi(X^i(x^-) = -2^{-n/2} (x^-)^{n+\frac{1}{2}}) \qquad \text{for }n \in \mathbb Z + \frac{1}{2},\nonumber
\end{eqnarray}
one gets the algebra
\begin{eqnarray}
\mbox{} [ \hat L_m,\hat L_n ] &=& (m-n)\hat L_{m+n},\nonumber\\
\mbox{} [ \hat{L}_m,\hat{N}_n ] &=& ( -\frac{z-2}{z}(m+1) -n )\hat{N}_{m+n} \nonumber ,\\
\mbox{} [\hat L_m,\hat X^i_n] &=& (\frac{m}{z} -n + \frac{2-z}{2z}) \hat X^i_{m+n}, \label{Alg3}\\
 \mbox{} [\hat X^i_m,\hat X^j_n] &=& (m-n) \hat N_{m+n}\delta^{ij}, \nonumber\\
\mbox{} [M_{ij},\hat X^k_n] &=& -\delta^{ik} \hat X^j_n +\delta^{jk} \hat X^i_n  , \nonumber\\ 
\mbox{}  [ \hat N_m ,\hat  N_n ] &= & 0 ,\qquad   [\hat N_m,\hat X^i_n] = 0,\nonumber
\end{eqnarray}
which generalizes to arbitrary $z$ the Schr\"odinger-Virasoro algebra studied in \cite{Henkel:1993sg, Unterberger:2007hd} for $z=2$ and the one proposed in \cite{Alishahiha:2009nm}.

When $z \neq 2$, the exact Killing vectors are given by $M_{ij}$, 
\begin{eqnarray}
\hat L_0 =  (-{r \over z}, -x^-,{z-2 \over z} x^+,-x_1 /z,...,-x_d/z)& &  \hspace{1cm} \mbox{dilatation} ,\no \\
\hat L_{-1} = (0,-\sqrt{2},0,0,...,0) & &  \hspace{1cm} x^- \,  \mbox{  translation} \label{exact1},\no \\
N_0 =  (0,0,1,0,...,0)  & &   \hspace{1cm} x^+ \, \mbox{translation}, \\
\hat X_{1/2}^i =  (0,0,2^{-1/4}x^i ,0,..,-2^{-1/4} x^-,..,0)  & &   \hspace{1cm} \mbox{boost}, \no\\
\hat X_{-1/2}^i  =  (0,0,0,0,.. , -2^{1/4},..,0)  & &   \hspace{1cm} x^i \, \mbox{translation}. \no
\end{eqnarray}
The Killing vector $\hat L_{-1}$ will be interpreted as the Hamiltonian and $\hat N_0$ as the particle number. For $z=2$, special conformal transformations are part of the symmetries. The corresponding generator $\hat L_1$ is given by
\begin{eqnarray}
\hat L_{+1}= (- 2^{-1/2} \, x^- \,r, - 2^{-1/2}(x^-)^2,{\overrightarrow{x}^2+r^2 \over 2} 2^{-1/2},-2^{-1/2} x^1  x^-,...,-2^{-1/2} x^i  x_d) \no \\
\mbox{special conformal transformation}  .\label{exact2}
\end{eqnarray}
In that case, the $\hat L_{-1}, \,\hat L_0, \, \hat L_1 , \, X^i_{1/2}, \, X^i_{-1/2}$ and $\hat N_0$ form the algebra denoted as $\mathfrak{sch}_2(d)$. 

This infinite-dimensional algebra is a natural generalization of the \sch algebra. However, the appearance of this algebra in the asymptotic Killing equations does not imply that it is actually realized, i.e. associated with finite, conserved, integrable and well represented charges in a phase space containing interesting solutions. We now turn our attention to this issue. 

\subsection{Realization of the asymptotic symmetries on a phase space for $z=2$  \label{knsection}}

This section is devoted to realize the asymptotic symmetry algebra on a phase space for $z=2$ and $d>0$ containing solutions of physical interest and such that the charges are finite, integrable, asymptotically conserved and well represented via a Dirac bracket. In section \ref{KNprephasespace}, we start the construction of  this phase space by considering a two-parameter family of black brane solutions and checking whether or not the charges associated with the candidate asymptotic Killing vectors \re{candidakv}  for $z=2$ give finite, integrable, conserved and well represented charges. Next we will turn in section \ref{z2fullphasespace}, to the construction of a restricted phase space by acting with finite diffeomorphisms associated with the asymptotic Killing vectors \re{candidakv} (that fulfill the above conditions on the pre-phase space) on the black branes in order to obtain a phase space that is invariant under the asymptotic symmetry algebra. 

\subsubsection{Black branes for the critical exponent $z=2$\label{KNprephasespace}}

Building on earlier work of \cite{Herzog:2008wg,Adams:2008wt,Maldacena:2008wh}, the authors of \cite{Kovtun:2008qy} constructed for any dimension a class of black hole solutions which asymptotes to \re{Metric} for $z=2$:
\beq
    && ds^2 = r^2 h^{-\frac{d}{d+1}} \left(
    \left[ \frac{(f{-}1)^2}{4(h{-}1)} - f \right] r^2 dx^{-}{}^2 + 
    (1{+}f) dx^{+}dx^{-} + \frac{h{-}1}{r^2}dx^{+}{}^2 \right) \no \\ 
    & & \hspace{1cm} + \, 
    h^{\frac{1}{d+1}}
    \left( r^2 dx^i dx^i + \frac{dr^2}{r^2 f} \right),\ \ \ \ \label{KNblackholes} \\
    && A = \frac{1{+}f}{2h}r^2 dx^{-} - \frac{1{-}h}{h}dx^{+}\,, \label{AKN}\\
    && \phi = -\frac12 \ln h\,, \label{phiKN}
\eeq
where
$ h(r) = 1+{\beta^2r_0^{d+2}}/{r^{d}}$ and $ f(r) = 1-{r_0^{d+2}}/{r^{d+2}}$, $\beta$ is an arbitrary parameter, 
and the horizon is located at $r=r_0$. 
The metric  \re{KNblackholes}  and matter fields \re{AKN}  and  \re{phiKN} are solution of the following Einstein gravity action coupled to a dilaton and a massive vector field,  
\beq
 S &=& \frac{1}{16\pi G_{d+3}}
   \int\!\! d^{d+3}x\, \sqrt{-g}
\Large[ R -
   \frac{a}{2}(\partial_\mu\phi)(\partial^\mu\phi) -
   \frac{1}{4} e^{-a\phi}F_{\mu\nu}F^{\mu\nu}  \nonumber \\
&& - \frac{m^2}{2}A_\mu A^\mu - V(\phi)
  \Large] \,,  \label{actionKN} 
\eeq
where $G_{d+3}$ is the $(d{+}3)$-dimensional Newton's constant,
the scalar potential is given by 
$ V(\phi) = (\Lambda{+}\Lambda')e^{a\phi} + (\Lambda{-}\Lambda')e^{b\phi}$,
and the coefficients are 
$$
   \Lambda=-\frac{1}{2}(d{+}1)(d{+}2) \,,\quad
   \Lambda'=\frac{1}{2}(d{+}2)(d{+}3) \,,\quad
   m^2=2(d{+}2) \,,\quad 
   a=(d{+}2)b=2\frac{d{+}2}{d{+}1} \,.
$$
In order to be able to interpret the solution as a gravity dual to a finite temperature non-relativistic system, we are required to identify the $x^+$ coordinate as $x^+ \sim x^+ + 2\pi x^+_0$. The particle number $\mathcal N$ associated with $\d_+$ has then discrete values. Using the methods described in Appendix \ref{method}, the charge $(D-2)$-forms associated with exact symmetries (evaluated at constant $x^-$ and at any finite $r$) are found to be integrable in the phase space\footnote{This phase space is only a preliminary one. We should still consider all finite diffeomorphisms associated with the asymptotic Killing vectors to built the entire phase space in order for it to be invariant under the action of asymptotic symmetries.} parameterized by $\beta$ and $r_0$. We denote the set of fields given in \re{KNblackholes}-\re{phiKN} by $\Phi(\beta, r_0)$, 
\beq   \Phi(\beta, r_0) := \{ g_{\mu\nu}(\beta,r_0) , A_\mu(\beta,r_0), \phi(\beta,r_0)\}  .\eeq 
Setting the charges of the background $\bar \Phi = \Phi(0,+\infty)$ to zero by convention, the final expressions for the conserved exact charges are given by
\begin{eqnarray}
\mathcal N &\equiv& \mathcal Q_{-\d_+} = \frac{D-1}{16 \pi G_{d+3}} \frac{\beta^2 L^{D-2}}{r_0^{D-1}}(2 \pi x_0^+) \text{Vol}_d, \label{valueN}\\ 
\mathcal H &\equiv& \mathcal Q_{\d_-} =  \frac{D-3}{32 \pi G_{d+3}} \frac{L^{D-2}}{r_0^{D-1}}(2 \pi x_0^+) \text{Vol}_d, \\ 
\mathcal P_i &\equiv& \mathcal Q_{\d_i}=0, \quad \mathcal M_{ij} = 0, \label{otherch}
\end{eqnarray}
where $\text{Vol}_d = \int d^d x$ is the transverse volume and $D = d+3$. The Hamiltonian $\cH$ and particle number $\cN$ are finite provided we consider a finite volume $\text{Vol}_d$, \ie  we introduce a `box' in the $x^i$-space to regulate the charges. These expressions \re{valueN}-\re{otherch}   have been obtained using a Mathematica code\footnote{The Mathematica code can be downloaded from the homepage of G.C.} implementing the formulae for the charges in Appendix \ref{method} for $d=1,2,3$. The expression for general $d$ has been guessed by matching that in lower dimensions, but given the simplicity of the final expression, the result is expected to be valid for any $d$. The Hamiltonian is identical to the one of the anti-de Sitter black brane as expected from the Null Melvin Twist procedure \cite{Maldacena:2008wh,Adams:2008wt,Herzog:2008wg}.

If one plans to construct a phase space containing the black brane solutions \eqref{KNblackholes}, a necessary (but not sufficient) condition that any asymptotic symmetry $\xi_{as}$ of that phase space \emph{should} obey is that the charge $D-2$ form $\delta \mathcal Q_{\xi_{as}} = \int k_{\xi_{as}}[\delta_{\beta,r_0} \Phi(\beta,r_0) ; \Phi(\beta,r_0)]$ evaluated on $\Phi(\beta,r_0)$ for small perturbations of $r_0$ and $\beta$ should be finite, integrable and conserved. 
For a general candidate asymptotic Killing vector \eqref{candidakv}, we can show that the charge is indeed finite (if we introduce a box) and integrable. Computing the charge $\mathcal Q_{\xi_{as}}[\Phi(\beta,r_0) ; \bar \Phi] = \int_{\bar \Phi}^{\Phi(\beta,r_0)} \delta \mathcal Q_{\xi_{as}}$ of the solution $\Phi(\beta,r_0)$ with respect to the background $\bar \Phi$ \eqref{Metric}, we get the result
\begin{eqnarray}
\mathcal Q_{\xi_{as}}[\Phi(\beta,r_0) ; \bar \Phi] =   L(x^-) \mathcal H - N(x^-) \mathcal N + \frac{ \mathcal N }{\text{Vol}_d}\int d^d x (\vec{x} . \vec{X}'(x^-) + \frac{1}{4} \vec{x}^2 L''(x^-)) \,. 
\label{chargesr0beta} \end{eqnarray}
In particular, we get 
\begin{eqnarray}
\mathcal D &=& \mathcal Q_{2 \hat L_{0}} = -2 x^- \mathcal H , \\ 
\mathcal C &=& \mathcal Q_{-\sqrt{2} \hat L_{1}} = (x^-)^2 \mathcal H + \frac{\mathcal N}{\text{Vol}_d} \int d^d x  \frac{1}{2} \vec{x}^2  , \\ 
\mathcal K_i &=& \mathcal Q_{-2^{1/4} \hat X^i_{1/2}} =  \frac{\mathcal N}{\text{Vol}_d} \int d^d x x^i   .
\end{eqnarray}
The dilatations and special conformal transformations are explicitly $x^-$-dependent. They are therefore not explicitly conserved in time. We will however go back to the issue of conservation after having introduced the Dirac bracket of charges.

Let us now study if the above charges represent the algebra \re{Alg3} (with $z=2$). We should be careful to the fact that to have finite charges, we need to introduce a regulator, i.e. a finite box of integration $\int d^d x$. An important point is that the box is \emph{not} invariant under all the candidate asymptotic Killing vectors \re{candidakv} with a non-vanishing spatial component $\xi^i$, $i=1 \dots d$. Since the domain of integration is part of the data determining how to compute the charges, this could mean that the candidate asymptotic Killing vector modifying the location of the box should be removed from the asymptotic algebra. However, the regulator resembles more a technical obstacle than a physical limitation. Let us imagine that one could find a gravity dual to a NRCFT with fields (including the Hamiltonian density and particle number density) falling-off at spatial infinity $x^i \rightarrow \pm \infty$ instead of remaining constant. The system would be finitely extended and the charges associated with asymptotic symmetries would be defined. All these asymptotic symmetries would be interpreted in the dual picture as global symmetries of the boundary theory which are not preserved by particular solutions of that theory but which map a solution to another one with a transformed Hamiltonian and particle number density. 

Note also that we do not expect the algebra to be centrally extended. As shown in \cite{Unterberger:2007hd}, a central extension could only appear in the commutation relation of the Virasoro generators. Now, the central extension in three-dimensional AdS spacetime for example \cite{Brown:1986nw} is possible because the Virasoro modes are expanded in exponentials depending on an angular coordinate. The central term is given by the integral of some function of the Virasoro modes in this angular coordinate which leads to a Kroneker delta $\delta_{m+n,0}$ originating from the orthogonality relations of the exponentials. In our case, since the modes $\hat L_m$ are polynomials in $x^-$, variable that we do not integrate over, it is impossible to obtain a central term of the required form proportional to a Kroneker delta $\delta_{m+n,0}$. Therefore, the central term has to vanish. 

In order to define the action of symmetries on other generators including the ones which changes the shape of the box of integration, we will define the following Dirac bracket
\beq
\{ \cQ^{box}_{\xi_1}[\Phi ; \bar \Phi] , \cQ^{box}_{\xi_2}[\Phi ; \bar \Phi] \} :=  \delta^{\Phi}_{\xi_2} \cQ^{box}_{\xi_1}[\Phi ; \bar \Phi]+\delta^{box}_{\xi_2} \cQ^{box}_{\xi_1}[\Phi ; \bar \Phi]  , \label{modifiedPB}
\eeq
where the first term is the usual Dirac bracket involving the variation of the fields, while the second term
\beq
\delta^{box}_{\xi_2} \cQ^{box}_{\xi_1}[\Phi ; \bar \Phi] := \lim_{\eps \rightarrow 0} \frac{1}{\eps}\left( \cQ^{box(x-\eps\,  \xi_2)}_{\xi_1}[\Phi ; \bar \Phi]-\cQ^{box(x)}_{\xi_1}[\Phi ; \bar \Phi]  \right) \label{deltabox}
\eeq
accounts for the variation of the regulator. Here, we consider the box as some mapping of $S^1 \times S^d$ (with coordinates $y^+,y^i$) to the manifold parameterized by some functions $x^\mu(y^+,y^i)$. Using these definitions, one gets the expected results
\beq
\{ \mathcal X^i_m,\mathcal X^i_n \} &=& (m-n) \mathcal N_{m+n} \delta^{ij} ,\\
\{ \mathcal N_m,\mathcal L_n \}&=& m \,\mathcal  N_{m+n} , \label{aas}\\
\{ \mathcal X^i_m  ,\mathcal  L_n \} &=& (m-\frac n 2) \mathcal X^i_{m+n} \label{aas2},
\eeq
while the individual contributions in \eqref{modifiedPB} are not anti-symmetric under the exchange of $\xi_1$ and $\xi_2$ and thus do not make any sense by themselves. However, we also get the unexpected expressions
\beq
\{ \mathcal L_m , \mathcal L_n \} &=& (m-n) \mathcal  L_{m+n} - \frac{\cN }{\text{Vol}_d} \int d^d x \frac{\vec{x}^2}{4} \hat L_m \,\hat L_n''', \\
\{ \mathcal L_m,\mathcal N_n \} &=& 0,\label{anomaly} \\
\{ \mathcal L_m,\mathcal X^i_n \} &=& \frac{(m+1)(2m-2n-1)}{2(2m+2n+1)} \mathcal  X^i_{m+n} \label{anomaly2} , 
\eeq
which show that the Dirac bracket as defined in \eqref{modifiedPB} does not make sense in general since it is not anti-symmetric. However for the charges associated with the exact Killing vectors, it is easy to check that this Dirac bracket is well defined and is isomorphic to the algebra of exact symmetry generators. Note that we have to take into account the effect coming from the variation of the box for the exact charges to be correctly represented. The Dirac bracket could be ``anti-symmetrized'' by definition but it would not help since the average between e.g.  the correct right-hand side in \eqref{aas} and the incorrect right-hand side of \eqref{anomaly} would not be isomorphic to the algebra of generators.

Using the definition of the modified Dirac bracket, let us now notice that even though $\mathcal D$ and $\mathcal C$ are time dependent, their total time derivatives
\begin{eqnarray}
\frac{D}{D x^-}\mathcal D = \frac{\d }{\d x^-}\mathcal D + \{ \mathcal D,\mathcal H \} = -2\mathcal H+2\mathcal H = 0,\\
\frac{D}{D x^-}\mathcal C = \frac{\d }{\d x^-}\mathcal C + \{ \mathcal C,\mathcal H \} = 2 x^- \mathcal H + \mathcal D = 0
\end{eqnarray}
vanish as it should. Expanding in modes as in \eqref{modesLn}, we can also check that all Schr\"odinger charges $\mathcal L_n$, $n=-1,0,1$, $\mathcal N_0$, $\mathcal X^i_n$, $n=\pm\frac 1 2$ associated with asymptotic vectors with non-zero $L_n$, $N_n$ and $X_n^i$ respectively are totally conserved, 
\begin{eqnarray}
\frac{D}{D x^-}\mathcal L_n = 0 ,\qquad \frac{D}{D x^-}\mathcal N_n = 0 ,\qquad \frac{D}{D x^-}\mathcal X^i_n = 0.
\end{eqnarray}
This conservation property is familiar from the $AdS_3$ example in Einstein gravity \cite{Brown:1986nw} where even though the Virasoro charges depend explicitly on time, they are totally conserved because the symplectic flux at the boundary is zero. However, contrary to the $AdS_3$ example, the total derivative of the infinite-dimensional extension of those generators is not defined because the Dirac bracket is not defined.

At this point in the discussion, we could summarize as follows: only the exact symmetries of the background, \ie the Schr\"odinger algebra $\mathfrak{sch}_2(d)$,  are associated with well-defined charges on our pre-phase space provided that we introduce a regulator. 

\subsubsection{\label{z2fullphasespace}Restricted phase space for $z=2$}

The set of candidate asymptotic symmetries has been reduced to the set of exact Killing vectors. Let us now act with finite diffeomorphisms of parameter $p$ associated with any Schr\"odinger generator on the black brane solutions \eqref{KNblackholes}, and check that finiteness, conservation and integrability hold for these new solutions $\Phi[\beta,r_0,p] \equiv  (g[\beta,r_0,p],A[\beta,r_0,p],\phi[\beta,r_0,p])$ as well. 

We first focus on diffeomorphisms associated with the candidate asymptotic Killing vector \re{candidakv} with $L(x^-)=0$, we will specify the modes corresponding to exact Killing vectors only afterwards. The vector field 
\begin{eqnarray}
\xi_{as}(\vec{\mathfrak{X}}(x^-),\mathfrak{N}(x^-)) = (\mathfrak{N}(x^-) - \vec{x} . \vec{\mathfrak{X}}^\prime(x^-))\partial_+ + \mathfrak{X}^i(x^-)\d_- \label{diffeoXN}
\end{eqnarray}
generates the following active finite diffeomorphism of parameter $p$,
\begin{eqnarray}
x^i &\rightarrow & x^i + p \mathfrak{X}^i(x^-),\qquad  r \rightarrow r,\qquad  x^- \rightarrow x^- ,\no\\
 x^+ &\rightarrow & x^+ +p(\mathfrak{N}(x^-) - \vec{x} . \vec{\mathfrak{X}}^\prime(x^-))-\frac{p^2}{2}
 \vec{\mathfrak{X}}(x^-) . \vec{\mathfrak{X}'}(x^-).\label{diff0}
\end{eqnarray}
The integrability conditions  
\begin{eqnarray}
I \equiv \int_S \delta^{(2)}_{r_0,\beta,p} k_{ \xi_{as}(L(x^-),N(x^-))}[\delta^{(1)}_{r_0,\beta,p} \Phi(r_0,\beta,p) ; \Phi(r_0,\beta,p)] - ((1) \leftrightarrow (2) ) = 0
\end{eqnarray}
should hold for all asymptotic symmetries $\xi_{as}(L(x^-),N(x^-))$ (see eq. \eqref{candidakv}) of interest. 
We get that
\begin{eqnarray}
I = \frac{1}{\text{Vol}_d} \int d^dx  \left( \mathfrak{N}'(x^-) - (\vec{x} + p \vec{\mathfrak{X}}(x^-)). \vec{\mathfrak{X}}''(x^-) \right)\no \\
\times \left( \delta^{(1)}(\frac{\mathcal N}{2\pi x_0^+}) \delta^{(2)}p - [(1) \leftrightarrow (2)] \right) L(x^-),
\label{nonint1}
\end{eqnarray}
where $\mathcal N$ is the particle number depending on $\beta$ and $r_0$ given in \eqref{valueN}.  For the modes corresponding to exact symmetries, \ie ${\mathfrak N}=1$, $\mathfrak{X}^i=-2^{-1/4} x^-$ and $\mathfrak{X}^i=-2^{1/4} $, the integrability condition $I=0$ is fulfilled. \newline

Let us also compute the integrability condition for the diffeomorphisms associated with a non-zero $L(x^-)$.  We focus on a particular mode of $L(x^-)$ :  $\hat L_n(x^-)= -2^{-n/2}(x^-)^{n+1}$ and will specify to the exact modes only afterwards. The finite diffeomorphisms have the form
\begin{eqnarray}
x^- \rightarrow x^-(1-p n (x^-)^n)^{-1/n}, \qquad x^i &\rightarrow & x^i(1-p n (x^-)^n)^{-(n+1)/(2n)},\no\\  
r \rightarrow r(1-p n (x^-)^n)^{-(n+1)/(2n)},\qquad  x^+ &\rightarrow & x^+ -\frac{n+1}{4}\frac{r^2+\vec{x}^2}{x^-} \left((1-p n (x^-)^n)^{-1}-1\right),\nonumber \label{diff1}
\end{eqnarray}
and we obtain
\begin{eqnarray}
I = \frac{1}{\text{Vol}_d} \int d^dx  \left( n(n^2-1) \frac{\vec{x}^2}{4} (x^-)^{n-2} (1-p n (x^-)^n)^{-\frac{1+7n}{2n}} \right) \no\\
\times \left( \delta^{(1)}(\frac{\mathcal N}{2\pi x_0^+}) \delta^{(2)}p - ((1) \leftrightarrow (2)) \right) L(x^-). 
\label{nonint2}
\end{eqnarray}
This expression vanishes for $n=-1,0,1$. This fact is consistent with the expectation that all the exact symmetries of the background will belong to the asymptotic symmetry algebra. Remark that if we were able to define a modified Dirac bracket that represents correctly all the charges associated with $\xi_{as}$, we would conclude from expressions \eqref{nonint1}-\eqref{nonint2} that we have to fix the number of particles $\mathcal N = \text{constant}$ in order to get integrable charges. 

On the phase space constructed by acting with Schr\"odinger diffeomorphisms only, the conserved charges (which are complicated non-linear functions of the metric) are finite (if we introduce a `box'), conserved and well represented. The asymptotic symmetry algebra can be summarized as follows: 
\begin{itemize}
\item[(i)] strictly speaking, the asymptotic algebra is empty since our charges are either null or infinite, the phase space is therefore also empty;
\item[(ii)] if we introduce a box, the infinite charges are regulated. We need to restrict the asymptotic symmetry algebra to the exact symmetry algebra in order for the charges to be well represented. If we require the box to be invariant under the asymptotic symmetry algebra, we get as asymptotic algebra only the Hamiltonian $\hat L_{-1}$ and the particle number $\hat N_0$ (supplemented by the rotations $M_{ij}$ if we choose the box to be a sphere centered at the origin of the $x$-space);
\item[(iii)] if we allow the box to be acted upon by other generators, the asymptotic symmetries consist of all exact generators $\mathfrak{sch}_2(d)=\{\hat L_{-1},\hat L_0,\hat L_1, $ $ \hat N_0, \hat X_{-1/2}, \hat X_{1/2}, M_{ij} \}$.
\end{itemize}

The phase space constructed in the previous sections is extremely limited since it contains no bulk excitations. It would be interesting to define boundary conditions including at the same time bulk excitations and Schr\"odinger asymptotic symmetries.  However, given the non-linearities in the asymptotic region, such an analysis would be pretty tedious, see however \cite{Ross:2009ar}.

\subsection{Realization of the asymptotic symmetries on a phase space for $z>2$ : an example \label{MBH}}

The generic family of black brane solutions with $z\neq 2$ in any dimension is not known. We will therefore analyze the case $z>2$ by considering a particular case : $z=3$ in $D=5$. As for the $z=2$ case, we will first compute the charges for a family of black holes depending on two parameters and verify their conservation, finiteness (up to a regulator), integrability and representation through a Dirac bracket. Since the computations are analogous to the ones of the  $z=2$ case, the asymptotic algebra will \emph{a priori} not contain any infinite-dimensional extensions of the exact symmetry group of the background.  Next we turn to the construction of the entire phase space by acting with finite diffeomorphisms associated with the asymptotic Killing vectors.  

\subsubsection{Black holes with $z=3$ in $d=2$ \label{solM}}

For $z\neq 2$, we do not generically know black hole solutions that asymptote \re{Metric}. But nicely, for the particular case of $z=3$ in $d=2$, the following black hole metric\footnote{We thank M. Rangamani for sharing his unpublished notes on these solutions analogous to the ones of \cite{Hubeny:2005qu}.} 
\beq
ds^2&=& {1 \over r^2 f(r)} dr^2 - (dx^-)^2 ({f(r)\over r^6}-{r^2 r_+^4 \over 4 \beta^2} )  + dx^- dx^+ ({1+f(r) \over  r^2})+ r^2 r_+^4 \beta^2 (d x^+)^2 \no \\ &&+ { dx_1^2 + dx_2^2  \over r^2}
, \label{bhM} \eeq where $f(r)= 1 -{r_+^4 r^4} $ does asymptote \re{Metric}. It is a solution of the action 
\beq
 S &=& \frac{1}{2 \kappa^2}
   \int\!\! d^{5}x\, \sqrt{-g} ( e^{-2 \phi} (R - 2 \Lambda 
 -  \frac{1}{12} H_{\mu\nu\rho} H^{\mu\nu\rho}) - {1 \over 12} F_{\mu\nu\rho}F^{\mu\nu\rho}) \no \\
&& + {2 \over \kappa^2}\int B \wedge  F   \,,  \label{actionM}
 \eeq
 where $F = dC $ and $H = dB$. The metric  \re{bhM} is supported by the following cosmological constant and matter fields
 \beq
 B &=& ({1+f(r) \over r^4}\, dx^- + 2 r_+^4 \beta^2 \,  dx^+) \wedge dx_1 \, , \label{BM}\\
 C&=&  -2e^{-\phi} {f(r)\over r^4} \, dx^- \wedge dx_2
 , \label{CM} \\
  \Lambda &=& -10+\frac{4}{1+r_+^4 \beta^2}\, , \quad  e^\phi  =  { 1 \over \sqrt{1 + r_+^4 \beta^2}} \label{phiM} \, .\eeq
In the coordinates chosen, even though the metric asymptotically approaches the one of \eqref{Metric}, the dilaton and the field $C$ have different value at infinity for different values of $r_+^4 \beta^2$. The fields are therefore not strictly speaking asymptotic to the zero-temperature solution. This will make the analysis of asymptotic charges quite subtle. In order to describe a non-relativistic system with a discrete spectrum for the particle number, we should identify the $x^+$ coordinate as $x^+ \sim x^+ + 2\pi x^+_0$. Hence, any transformation which does not depend periodically on $x^+$ cannot exist. In particular, the dilatation and all Virasoro generators which are part of the candidate asymptotic Killing vectors \eqref{candidakv} cannot be part of the asymptotic symmetries, except the Hamiltonian for which $L'(x^-) = 0$ (in contrast to the $z=2$ case).

It turns out that the charge $D-2$-form associated with the generator $\d_-$ is not integrable in the phase space parameterized by $\beta$ and $r_+$. Therefore, the Hamiltonian cannot be associated with $\d_-$ following standard prescriptions. One way to define the Hamiltonian consists in multiplying the generator $\d_-$ by an ``integrating factor'' $f(r_+,\beta)$ chosen such that the resulting charge is integrable, see \cite{Barnich:2007bf}. One finds that $f(r_+,\beta)$ has to have the form
\beq
f(r_+,\beta) = r_+^2 \tilde f(r_+^4 + \beta^2 r_+^8)
\, . \eeq
 A natural choice for $\tilde f$ is to require that the integrating factors goes to 1 when $r_+$ goes to zero. The resulting unique factor is given by 
\beq
f(r_+,\beta) = \frac{1}{ \sqrt{1 + \beta^2 r_+^4}}
\, , \eeq
which is in fact the same expression as the dilaton which is non-trivial at infinity. The Hamitonian is then defined as
\beq 
\cH \equiv \frac{1}{f(r_+,\beta)} \cQ_{ f(r_+,\beta) \partial_-}.\label{presc}
\eeq
We will see that this is the correct prescription to obtain an isomorphism between the algebra of asymptotic symmetries and the Dirac bracket\footnote{In the treatment of \cite{Barnich:2005kq}, the integrating factor that was considered in order to define the energy was not compensated by an overall inverse integrating factor in front of the integrated charge. The current prescription would also be natural in that context, for it would reproduce the expectation that the energy of G\"odel black holes and black strings in pp-waves spacetimes are equal since those solutions are related by dualities \cite{Gimon:2003xk}.}. 

Setting the charges of the background to zero by convention, the final expressions for the charges associated with the vectors \re{candidakv} are given by
\beq 
\cH \equiv  f(r_+,\beta)^{-1} \cQ_{ f(r_+,\beta) \partial_-}&=& {r_+^4(1 + r_+^4 \beta^2)^{3/2} \over 16 \pi G}Ê\, (2 \pi x^+_0) \, \text{Vol}_d \, ,\no \\
\cN \equiv \cQ_{-\partial_+} &= & \frac{r_+^4 \beta^2 (1 + r_+^4 \beta^2 )^{3/2}}{4 \pi G} \, \, (2 \pi x^+_0) \, \text{Vol}_d \, ,\no\\
\cQ_{ \hat X^i_n}&= &{\cN\over \text{Vol}_d} \int d^dx  \, x^i  X_n^{i\prime}(x^-)\, ,\\
\cQ_{\hat N_n}&= & - N_n(x^-) \cN \, ,\no
\eeq
where
prime denotes derivative with respect to $x^-$. Note that the charge associated with translations $\mathcal X^i_{-1/2}$ and angular momentum are zero. Using the definition of the Dirac bracket \eqref{modifiedPB}, one obtains $\{ \mathcal H,\mathcal N_n \} = 0$, for all $n \in \mathbb Z$ and $\{ \mathcal H , \mathcal X_n^i\}= 0$, for all $n \in \mathbb Z+\frac{1}{2}$. We see that the isomorphism with the symmetry algebra \eqref{Alg3} holds only for the expected generators $\hat N_0$, $\hat X^i_{-1/2}$ and $\hat X^i_{1/2}$ while the representation of the infinite-dimensional generalizations of these generators breaks down, exactly as in the $z=2$ case. One can check that the remaining Dirac brackets have the expected commutation rules.

\subsubsection{Restricted phase space for  $z=3$, $d=2$}

We could act with the finite diffeomorphisms associated with the candidate asymptotic symmetries on the black holes \re{bhM} to construct a restricted phase space. According to the analysis done in the previous section, the candidate asymptotic symmetries are reduced to the Galilean algebra and the particle number $\xi_{cand} = \{ \hat H,\hat N, \hat X^i_{-1/2},\hat X^i_{1/2},\hat  M_{12} \}$. It is then straightforward to check that the family obtained by acting with the finite diffeomorphisms associated with these vectors on the black brane solutions is a good phase space, \ie is invariant under the Galilean algebra together with the particle number, and is such that all charges on the family are finite (up to the regulator), integrable, totally conserved and well represented via the generalized Dirac bracket \eqref{modifiedPB}.

\section{Conclusion and discussion  \label{conclusion}}

We have studied the representation of asymptotic charges in asymptotically Schr\"odinger spacetimes. While there exists a consistent infinite-dimensional algebra which extends the Schr\"odinger algebra in any dimension and for any dynamical exponent $z$, the charges associated with these generators have been shown not to obey a regular Dirac bracket algebra in the sense of Brown-Henneaux.

Our derivation proceeded by providing a Lagrangian method to derive the conserved charges of black branes in Schr\"odinger spacetimes. Since these branes are infinitely extended, they require a cut-off in each spatial direction. The regularized charges then depend on this spatial cutoff which is not invariant under the whole Schr\"odinger algebra. A Dirac bracket between two charges including the variation of the cut-off was defined and was shown to represent the asymptotic Schr\"odinger algebra of symmetries. Moreover, the Schr\"odinger asymptotic charges were shown to be conserved in the sense that the total derivative of the charges, including both the explicit time dependence and the commutator with the Hamitonian, was shown to be zero. However, none of the proposed generators in the infinite extension of this algebra appeared to have well-defined Dirac brackets on the restricted phase space of black branes, i.e. on the finite-temperature solutions. We thereby concluded that the infinite-dimensional extension is not part of any asymptotic symmetry algebra of a phase space containing these black branes. We can thus argue that non-relativistic systems having a gravity dual will contain fields forming representations of the Schr\"odinger group, and not the Schr\"odinger-Virasoro group.

Let us now discuss some extensions and directions for future developments. Our asymptotic analysis is identical if one considers the global coordinates for the Schr\"odinger metric obtained in \cite{Blau:2009gd} since the behavior of the metric only differs from the metric we studied by terms becoming subleading at the boundary. 
Also, since the charges are regulated using a box in all dual spatial directions, one could equivalently consider an infinitesimal box or, equivalently, charge densities and the same conclusions would apply. 

An interesting possibility comes from the Schr\"odinger spacetime with a spherical spatial boundary described in \cite{Yamada:2008if}. One could expect that the finite area spatial boundary would give finite charges without needing a regulator which introduced all the problems in the representation of the charges\footnote{We thank A. Adams for a discussion on that issue.}. Therefore, an infinite-dimensional extension in these backgrounds is not discarded. The boundary theory would however have to be defined on a sphere which is pretty usual from the condensed matter perspective. 

The canonical charges associated with the generators of time-translation, translation in the compact null direction and spatial translations were obtained straightforwardly for $z=2$. For $z=3$, however, a subtle manipulation of the conserved charge was necessary in order to define an integrable charge which is still associated with the canonical time and which still represent the algebra of asymptotic symmetries. General results on the equivalence of Hamiltonian and Lagrangian formalisms, and the unicity of the charges shows that identical results would be obtained in Hamiltonian framework if one also uses the  prescription \eqref{presc} we introduced for the integration in phase space. 

We also comment in Appendix \ref{Lifsec} on the relationship between our results and another class of gravitational backgrounds relevant to the non-relativistic AdS/CFT correspondence, namely the Lifshitz spacetimes \cite{Kachru:2008yh}. We show that a candidate infinite-dimensional extension of the Lifshitz symmetry can be defined. However, a regulator and a modified Dirac bracket should be defined. This can be argued to lead to the same problems as the ones encountered in the Schr\"odinger case.

Another approach to look at gravitational backgrounds dual to NRCFTs with Schr\"odinger invariance relies on the observation that, in non-relativistic systems, the number or mass operator usually appears as a central element between the translations and boosts instead of being a generator on its own\footnote{We thank A. Maloney for sharing his thoughts on these questions and for his suggestions.}. Since central extensions cannot appear in the bracket of exact symmetries, one idea would be to look at spacetimes which do not admit the full Schr\"odinger algebra as an exact symmetry group, but instead realize it as its asymptotic symmetry group. One natural question is to ask if the Lifshitz spacetimes can realize such a scenario since they admit translations but not boosts as exact symmetries. However, the exact statement is that the central elements can appear only in the Dirac bracket between two asymptotic symmetries \cite{Brown:1986ed}. This is easily seen by using the anti-symmetry of the central charge,
\begin{eqnarray}
\cK_{P_i,K_j} = \int_{S^\infty} k_{P_i}[\mathcal L_{K_i}\bar\Phi ; \bar\Phi] = - \int_{S^\infty} k_{K_j}[\mathcal L_{P_i}\bar\Phi ; \bar\Phi] 
\end{eqnarray}
between the translations and rotations, where $\bar \Phi$ are the fields of the background including the metric. Since the Lifshitz spacetime is translation-invariant, it is not appropriate to realize that idea. The only way the number operator could appear as central element would be to consider a gravity background where both translations and Galilean boosts would be realized as asymptotic isometries.

In place of Schr\"odinger algebras, NRCFT can be based on Galilean conformal algebras \cite{Henkel:1997zz,Lukierski:2005xy}. The proposal of gauge/gravity correspondences based on Galilean conformal algebras has been developed so far using the Newton-Cartan formalism (see e.g. \cite{Duval:2009vt} and references therein). It has been argued recently in \cite{Bagchi:2009my} that infinite-dimensional extensions of the asymptotic symmetry group could occur in that context as well. Unfortunately, our charge analysis does not extend straightforwardly to this case since the lack of a regular metric in the bulk would prevent one to use covariant phase space methods to define the conserved charges of the theory to infirm or confirm the proposal.

In this paper we focused on spacetimes of dimensions strictly greater than $3$ which are conjectured to be dual to field theories living in a positive number of spatial dimensions. However, from the classical asymptotic analysis of AdS spaces of  \cite{Brown:1986nw, Henneaux:1985ey, Henneaux:1985tv}, it is expected that the three-dimensional background will exhibit specific features with respect to its higher-dimensional counterparts. One can show it is indeed the case: as for $AdS_3$, the asymptotic symmetry algebra becomes infinite-dimensional with completely well-defined charges satisfying a Virasoro algebra. Those results will be presented elsewhere.

\section*{Acknowledgments}

We would like to thank Allan Adams, Nicolay Bobev, Gaston Giribet, Sean Hartnoll, 
Mokhtar Hassaine, Marc Henneaux, Petr Ho\v{r}ava, Veronika Hubeny, Josh Lapan, Alex Maloney, Don Marolf, Charles Melby-Thompson, Michael Mulligan, Mukund Rangamani, Sakura Schafer-Nameki, Philippe Spindel, Andy Strominger and J\'er\'emie Unterberger for fruitful discussions. 
The work of GC is supported in part by the US National Science Foundation under Grant No.~PHY05-55669, and by funds from the University of California.
The work of SdB and SD are funded by the European Commission though the grants PIOF-GA-2008-220338 and PIOF-GA-2008-219950  (Home institution:
Universit\'e Libre de Bruxelles, Service de Physique Th\'eorique et
Math\'ematique, Campus de la Plaine, B-1050 Bruxelles, Belgium). The work of KY was supported in part by the National Science Foundation 
under Grant No. PHY05-51164 
and by the Grant-in-Aid for the Global COE Program "The Next Generation of Physics, Spun from Universality and Emergence" from the Ministry of Education, Culture, Sports, Science and Technology (MEXT) of Japan.

\appendix

\section{Method to compute conserved charges \label{method}}

In this appendix, we will briefly review the formalism of \cite{Barnich:2001jy,Barnich:2003xg,Barnich:2007bf} to  
compute conserved or asymptotically conserved charges. We will present the method for gravity in $D 
$ dimensions coupled to one $p$-form and then provide the relevant definitions for a more general Lagrangian including multiple $p$-forms, scalar fields as well as $U(1)$ and gravitational Chern-Simons terms.

\subsection{General definitions illustrated on an example}
\label{method1}

Let us explain how conserved charges are defined on an example : the  
Einstein--$p$-form system in $D$ dimensions with the following action,
\begin{equation}
I = \frac{1}{16 \pi G} \int \, d^Dx \,\left[ \sqrt{-g}\left( R
%+ \frac{2}{l^2}
- \frac{1}{2} \star \mathbf F \wedge \mathbf  F\right)
%- \frac{\alpha}{2}\eps^{\mu\nu\rho} A_\mu F_{\nu\rho}
\right],\label{action}
\end{equation}
where $\mathbf F = d \mathbf A$.
The gauge parameters of the theory
$(\xi, \mathbf\Lambda)$, where $\xi$ generates infinitesimal  
diffeomorphisms and ${\mathbf\Lambda}$ is the parameter of $U(1)$  
gauge transformations are endowed with the Lie algebra structure
\begin{equation}
[(\xi, \mathbf\Lambda),(\xi^\prime, \mathbf\Lambda ^\prime)]_{G} =
([\xi,\xi^\prime],[\mathbf\Lambda, \mathbf\Lambda ^\prime]),\label 
{eq:Lie}
\end{equation}
where the $[\xi,\xi^\prime]$ is the Lie bracket and
$[\mathbf\Lambda, \mathbf\Lambda ^\prime]\, \equiv \cL_\xi \mathbf 
\Lambda ^\prime -
\cL_{\xi^\prime} \mathbf\Lambda $. We will denote for compactness the  
fields as $\phi \equiv (g_{\mu\nu}, \mathbf A)$ and the gauge  
parameters as $f = (\xi^\mu, \mathbf\Lambda)$. For a given field $\phi 
$, the gauge parameters $f$ satisfying
\begin{equation}
\cL_\xi  g_{\mu\nu} \approx 0, \qquad \cL_\xi
\mathbf A + d \mathbf\Lambda \approx 0, \label{eq:red}
\end{equation}
where $\approx$ is the on-shell equality, will be called the exact  
symmetry parameters of the field configuration $\phi$. Parameters $(\xi, \mathbf\Lambda)  
\approx 0$ are called trivial symmetry parameters. The set of gauge parameters which satisfy the equations \eqref{eq:red} in an asymptotic region, i.e. such that in some large radius $r$ limit the equations are satisfied at leading order, and which form a Lie algebra, will be called 'candidate asymptotic symmetries'. The concept of (truely) asymptotic symmetries are defined as a subset of those which are associated to finite, conserved and integrable charges, see the next definitions.

\medskip

It exists a canonical algorithm to construct a spacetime $D-2$ form
\begin{eqnarray}
\mathbf k_{f} [\delta \phi ; \phi ], \label{oneform}
\end{eqnarray}
which is also a one-form in field space (because the expression is linear in $\delta \phi$ and its derivatives) such that the following properties hold : 
\begin{itemize}
\item The conserved quantity associated with any exact symmetry parameter $f$ that provides the difference of charge between the solution $\phi$ and the solution $\phi + \delta \phi$ where $\delta \phi$ obeys the linearized equations of motion is given by 
\beq
\delta Q_{f} := \oint_S \mathbf k_{ f} [\delta \phi ; \phi ]  \label{infcharge}
\eeq
and only depend on the homology class of the $D-2$ surface $S$. As a consequence, the conserved charge \eqref{infcharge} is finite and time-independent. One can further show that the conserved charge is unique, i.e. there is a one-to-one correspondence mapping a couple of symmetry parameters and a surface of given homology class and conserved charges \eqref{infcharge} \cite{Barnich:1994db}.

\item The quantity associated with a candidate asymptotic symmetry parameter $f$ that provides the difference of charge between the solution $\phi$ and the solution $\phi + \delta \phi$ where $\delta \phi$ obeys the linearized equations of motion is given by 
\beq
\delta Q_{f} := \text{lim}_{r \rightarrow \infty} \oint_{S^r} \mathbf k_{ f} [\delta \phi ; \phi ] . \label{infchargeasympt}
\eeq
This quantity can be infinite and/or not conserved depending on the choice of boundary conditions obeyed by $\phi$ and $\delta \phi$. Given a definition of phase space, one has to discard any candidate asymptotic symmetry which violates the conditions of finiteness and conservation of the charges. 

\item The form \eqref{oneform} is constructed out of the equations of motion and therefore does not depend on boundary terms that may be added to the Lagrangian. Moreover, the form is a linear functional of the equations of motion, and so, of the Lagrangian. One can therefore construct this form by summing up the individual contributions from the different pieces of the Lagrangian. 
\end{itemize}
Additional properties of the charge form \eqref{oneform} are discussed in \cite{Barnich:2004uw,Barnich:2006av}. In the case of the Lagrangian \eqref{action}, one gets
\begin{eqnarray}
\mathbf k_{ \xi,\mathbf\Lambda} [\delta \phi ; \phi ] &=& 
\mathbf k^{g}_{ \xi}[\delta g;g]  +  k^{\mathbf A}_{ \xi,\mathbf 
\Lambda}[\delta \phi ; \phi ] , \label{k_tot}
\end{eqnarray}
where the gravitational contribution to the charge  form is given by \cite{Abbott:1981ff,Barnich:2001jy}
\begin{eqnarray}
\mathbf k^{g}_{ \xi}[\delta g;g] &=& -\delta \mathbf Q^g_{ \xi} -i_{\xi}\mathbf \Theta^g[\delta g] -\mathbf
E^g_\cL[\cL_\xi g, \delta g],\label{grav_contrib}
\end{eqnarray}
where
\begin{eqnarray}
\mathbf Q^g_{\xi}&=& \star \Big( \half (D_\mu\xi_\nu-D_\nu\xi_\mu)
dx^\mu \wedge dx^\nu \Big),\label{Komar_term} \\
\mathbf \Theta^g[\delta g]&=&\star \Big(  (D^\sigma \delta
g_{\mu\sigma}-
g^{\alpha\beta} D_\mu \delta g_{\alpha\beta})\,dx^\mu\Big),\\
\mathbf E^g_\cL[\delta_2 g, \delta_1 g] &=& \star \Big( \half \delta_1
g_{\mu\alpha} g^{\alpha\beta }\delta_2 g_{\beta\nu} dx^\mu \wedge
dx^\nu \big).
\end{eqnarray}
The term \re{Komar_term}  is the Komar $D-2$ form and the  
supplementary term, $E^g_\cL$, with respect to the Iyer-Wald
form~\cite{Iyer:1994ys} vanishes for Killing vectors but may be relevant  
for asymptotic symmetries. In \eqref{k_tot}, we define $\delta$ as an operator acting on the fields $\phi$ but not on $\xi$. 
 The $p$-form contribution to the charge form is given by \cite{Compere:2007vx}
\begin{equation}
\mathbf k^{\mathbf A}_{ \xi,\mathbf\Lambda}[\delta \phi ; \phi ]=- 
\delta \mathbf
Q^{\mathbf A}_{\xi,\mathbf\Lambda} + i_\xi \mathbf
\Theta_{\mathbf A}-\mathbf E^{\mathbf A}_\cL[\cL_\xi \mathbf A+d
\mathbf\Lambda,\delta \mathbf A] \label{Bcharge}
\end{equation}
with
\begin{eqnarray}
& &Q^{\mathbf A}_{\xi,\mathbf\Lambda}  =  (i_\xi \mathbf A + \mathbf\Lambda)   
\wedge \star \mathbf F \label{QA} ,\qquad \mathbf \Theta^{\mathbf A} = \delta \mathbf A
\wedge \star \mathbf F,\label{ThetaA}\\
& &\mathbf E^{\mathbf A}_\cL[\delta_2 \mathbf A,\delta_1 \mathbf A] =  \star \big(
\half \frac{1}{(p-1)!}\delta_1
\mathbf A_{\mu\alpha_1\cdots \alpha_{p-1}} \delta_2 \mathbf A_{\nu}^{\;\,\,\alpha_1\cdots \alpha_{p-1}} dx^\mu\wedge
dx^\nu \big).
\end{eqnarray}

\medskip

\noindent The set of fields $\phi$, $\delta \phi$ and gauge parameters  
$(\xi,\mathbf\Lambda)$ that satisfies the conditions
\begin{eqnarray}
\oint_{S} \delta_1 k_{f}[\delta_2 \phi,\phi] - (1 \leftrightarrow 2)= 0,  
\label{cond_int} \\
\oint_{S} \mathbf E_\cL[\delta_1 \phi, \delta_2 \phi] - (1 \leftrightarrow 2)  = 0,\label{cond_alg}
\end{eqnarray} 
define a space of fields and parameters which we denote as the integrable space $\cI$. In this space, we define
the charges difference between the reference field $\bar \phi$ and the field $\phi$
associated with $f= (\xi,\mathbf\Lambda)$ as
\begin{equation}
\cQ_{(\xi,\mathbf\Lambda)}[\phi,\bar \phi] = \oint_{S} \int_ 
\gamma k_{(\xi,\mathbf\Lambda)}[\delta \phi,\phi] + \cN_{(\xi,\mathbf 
\Lambda)}[\bar \phi],
\end{equation}
where $\gamma$ is a path in field space contained in $\cI$ and $\cN_ 
{(\xi,\mathbf\Lambda)}[\bar \phi]$ is an arbitrary normalization  
constant. The condition~\eqref{cond_int} ensures that the charge is  
independent on smooth deformations of the path $\gamma$. The  
condition \re{cond_alg} is a technical assumption needed for the representation theorem, see below. A candidate asymptotic symmetry $f[\phi]$ will be called an asymptotic symmetry of a given phase space at $\phi$ if the conserved charges associated to $f[\phi]$ around $\phi$ are all finite, conserved and integrable.

\medskip

Let us denote as $\cA$ the largest algebra of asymptotic symmetries $f[\phi] = (\xi[g,\mathbf A],\mathbf \Lambda[g,\mathbf A])$ such that for each 
field $\phi$ in the phase space the set of parameters $f[\phi]$ form a closed Lie algebra under the bracket  
defined in \eqref{eq:Lie} and such that all these algebras are  
isomorphic. Using the conditions~\eqref{cond_int}-\eqref{cond_alg}, one can then show that for any solutions $\bar \phi$ and $\phi$ in the  
integrable space, and for any $(\xi,\lambda)$, $(\xi^\prime, \lambda^ 
\prime)$ in $\cA$, the Dirac bracket defined by
\begin{equation}
\left\{ \cQ_{(\xi,\mathbf\Lambda)}[\phi,\bar \phi],
\cQ_{(\xi^\prime,\mathbf\Lambda^\prime)}[\phi,\bar \phi] \right\} \equiv
\oint_{S^\infty} k_{(\xi, \mathbf\Lambda)}[(\cL_{\xi^\prime}
g_{\mu\nu},\cL_{\xi^\prime} \mathbf A + d \mathbf\Lambda^\prime);\phi] \label{poissonbracket}
\end{equation}
can be written as
\begin{equation}
\left\{ \cQ_{(\xi,\lambda)}[\phi,\bar \phi],
\cQ_{(\xi^\prime,\mathbf\Lambda^\prime)}[\phi,\bar \phi] \right\} =
\cQ_{[(\xi,\mathbf\Lambda),(\xi^\prime,\mathbf\Lambda^\prime)]_{G}} 
[\phi,\bar \phi] -
\cN_{[(\xi,\mathbf\Lambda),(\xi^\prime,\mathbf\Lambda^\prime)]_{G}} 
[\bar \phi] + \cK_{(\xi,\mathbf\Lambda),(\xi^\prime,\mathbf\Lambda^ 
\prime)}[\bar \phi],\label{formula}
\end{equation}
where \begin{eqnarray}
\cK_{(\xi,\mathbf\Lambda),(\xi^\prime,\mathbf\Lambda^\prime)}[\bar  
\phi] = \int_{S^\infty} k_{(\xi, \mathbf\Lambda)}[(\cL_{\xi^\prime}  
\bar g_{\mu\nu},\cL_{\xi^\prime} \bar{\mathbf A} +
d \mathbf\Lambda^\prime);\bar \phi] \label{eq:cc}
\end{eqnarray}
is a central extension which is considered as trivial if it  
can be reabsorbed in the normalization of the charges $\cN_{[(\xi, 
\mathbf\Lambda),(\xi^\prime,\mathbf\Lambda^\prime)]_{G}}[\bar \phi]$.

\subsection{Charge form for a more general Lagrangian}

For a general action with $r$ scalar fields $\overrightarrow{\chi} = 
\{ \chi_1, ... \chi_r \}$ and any number of $p$-form fields,
\begin{equation}
I = \frac{1}{16 \pi G} \int \, \left( R \, \star {\oneone}
- \frac{1}{2}  \star d  \overrightarrow{\chi} \wedge d \overrightarrow 
{\chi}
-  \frac{1}{2} \sum_a e^{-\overrightarrow{\alpha_a}. \overrightarrow  
\chi } \star \mathbf F^a \wedge \mathbf  F^a \right)
,\label{gaction}
\end{equation}
the charge form is given in terms of the building blocks defined in section \ref{method1} as
\beq
\mathbf k_{ \xi,\mathbf\Lambda^{ a}} [\delta \phi ; \phi ] &=& 
\mathbf k^{g}_{ \xi}[\delta g;g]  +  \sum_a e^{-\overrightarrow 
{\alpha_a}. \overrightarrow \chi} k^{\mathbf A^a}_{ \xi,\mathbf 
\Lambda^{a}}[\delta \phi ; \phi ] + \sum_i \mathbf k^{\chi^i}_{ \xi} 
[\delta \phi;\phi]  \nonumber \\
&& + \,\sum_{a} \mathbf k^{\mathbf A_a \, suppl}_ 
{\xi, \mathbf \Lambda_a}[\delta \phi;\phi]   , \label{k_totg}
\eeq
where
\beq
\mathbf k^{\chi^i}_{ \xi}[\delta \phi;\phi] &=& i_\xi \big(\star (d  
\chi^i \delta \chi^i  ) \big)  \hspace{1cm} \mbox{no sum over } \, i\ , \\
\mathbf k^{\mathbf A_a \, suppl}_{\xi, \mathbf \Lambda_a}[\delta \phi; 
\phi] &= &  \delta \overrightarrow{\chi}. \overrightarrow{\alpha} e^{- 
\overrightarrow{\alpha}_a . \overrightarrow{\chi}} {\cal Q}^{\mathbf A_a}_ 
{\xi , \mathbf\Lambda_a} .
\eeq
The last contribution can be understood by the fact that the charge form of the $p$-forms will have an expression similar to \eqref{Bcharge} with a Komar term ${\cal Q}^{\mathbf A^a}_{\xi , \mathbf\Lambda^a}$ including the factor   $e^{-\overrightarrow{\alpha_a}. \overrightarrow \chi }$.

In section \ref{solM}, we compute charges for a  
solution of a five dimensional theory with a Chern-Simons term \re{actionM}  of  
the following form
\beq
I_{CS} = \mathbf B \wedge d \mathbf C. \eeq
One can compute the corresponding contribution to the charge form and one gets 
\beq
k^{CS} [\delta \phi;\phi] = \delta \mathbf B \wedge i_\xi \mathbf C -  
i_\xi \mathbf B \wedge \delta \mathbf C \, .
\eeq
Note also that in the string frame $I = \frac{1}{16 \pi G} \int \,  
\left(  e^{- 2 \chi} R \, \star {\oneone} - {1 \over 2} \star d \chi  
\wedge d \chi + ... \right)  $, the gravitational contribution to the charge form is modified to 
\beq
\mathbf k^{g \, string frame}_{ \xi}[\delta \phi;\phi]  = e^{- 2 \chi} 
\, \mathbf k^{g}_{ \xi}[\delta \phi;\phi]- \delta (e^{- 2 \chi})  
{\cal Q}^g_\xi + \text{terms}(\partial \chi) \, ,
\eeq
where the last terms are proportional to at least one derivative of the dilaton and thus vanish if the dilaton is constant. They play no role for the solutions of interest in this paper.

\section{Candidate asymptotic symmetries for Lifshitz spacetimes}
\label{Lifsec}

Gravity duals to non-relativistic systems governed by Lifshitz
symmetry have also been considered \cite{Kachru:2008yh}. The
zero-temperature background 
\beq
ds^2 = {dr^2 \over r^2} - r^{-2z} dt^2 + r^{-2} dx^i dx^i \hspace 
{1cm} (i = 1,..., d) \,  \eeq
can be described formally as the Kaluza-Klein reduction along the null direction $x^+$ of the background \eqref{Metric}. The kinematics analysis of these spacetimes will therefore be very similar to the one performed in the main text. However, since the theory describing Lifshitz spacetimes is different than Schr\"odinger spacetimes, the analyses of conserved charges will be different. The lack of a Null Melvin Twist procedure will also prevent one to use correspondences with  AdS to derive the conserved charges. Since the only black hole solutions known so far are numerical \cite{Danielsson:2009gi,Bertoldi:2009vn,Bertoldi:2009dt}, we will not attempt to construct an analytical phase space in this paper and we will limit our discussion to kinematical aspects of the asymptotic symmetries.

For simplicity, let us focus on the $d=2$ case. We solved the asymptotic Killing equations up to certain convenient orders and obtained the  
following candidate asymptotic Killing vectors,
\beq
\xi_{asym} &=&   {r\over z}  L'(t)  \partial_r + L(t)  
\partial_t + (X^1(t) + x^2 M + {x^1 \over z} L'(t)) \partial_{x^1} \no \\  
& &+ (X^2(t)-x^1 M + {x^2\over z} L'(t)) \partial_{x^2}\, .
\eeq
The exact symmetries are recovered when $L''(t)=0$, $X^{1\, \prime}(t) 
=0$ and $X^{2\, \prime}(t)=0$. The Hamiltonian corresponds to $L(t) = 
1$, the dilations to $L(t) = 2 t$,  the $x^1$-translations to $X^1(t) 
=1$, the $x^2$-translations to $X^2(t)=1$ and the rotations to $M$. 
Defining the generators
\begin{eqnarray}
L_n  &=& \xi_{asym}( L(t)= -2^{-n/2}t^{n+1}) \,  \, \, \, \text{for }n  
\in \mathbb Z, \\
X^i_n &=&  \xi_{asym}(X^i(t)= -2^{-n/2}t^{n+1/2})\, \,   \,  \, \, \, \text{for }n  
\in \mathbb Z + \frac{1}{2},
\end{eqnarray}
we obtain the infinite dimensional algebra
\beq
\, i[ L_m , L_n] &=& (m-n) L_{m+n} \, \no , \\
\, i[L_m, X^i_n] &=&   ({m  \over z } - n + \frac{2-z}{2z}) X^i_{m+n}\, , \label{Lialg} \\
\, [X^i_m, X^j_n] &=& 0 \, . \no \eeq
The asymptotic symmetry algebra of Lifshitz spaces is a truncation of the Schr\"odinger algebra $\mathfrak{sch}_z(d)$ \eqref{Alg3}. The generalization of these candidate asymptotic symmetries to any dimensions is straightforward. It is amusing to observe that, for $d=1$ and $z=2$, \re{Lialg} is precisely the symmetry algebra of the Burgers equation driven by an external force relevant in turbulence theory (see e.g. \cite{Ivashkevich:1996et}). On the other hand, the two-dimensional metric ($d=0$) is also a solution of Einstein-Maxwell theory with negative cosmological constant, like $AdS_2$. It might therefore be interesting to see whether the corresponding asymptotic algebras admit central extensions, in the spirit of  \cite{Brown:1986nw, Hartman:2008dq}.

The realization of these symmetries on a phase space will lead to the same kind of difficulties we encountered for the Schr\"odinger case.  
Indeed, in order to compute the charges, we will have to integrate on the $x^i 
$-plane and therefore we will obtain infinite results. The infinite charges  
could then be regulated by introducing a `box'. The Dirac bracket will have to be modified in order to accomodate the action of generators on the regulator.  
We therefore expect that the infinite dimensional algebra will not be realized if we follow the same strategy as the one presented in this paper for  
the Schr\"odinger case. 

\providecommand{\href}[2]{#2}\begingroup\raggedright\endgroup

\end{document}